\documentclass[10pt,conference]{IEEEtran}

\usepackage[english]{babel}

\usepackage{amsmath}
\usepackage{graphicx}
\usepackage{soul}
\usepackage{xspace}
\usepackage{cleveref}
\usepackage{cite}
\usepackage{float}
\usepackage{url}

\newcommand{\tool}{Bus Factor Explorer\xspace}

\newcommand{\datasetSize}{935\xspace}

\title{Bus Factor Explorer}

\author{
\IEEEauthorblockN{Egor Klimov}
\IEEEauthorblockA{JetBrains Research\\Serbia\\
egor.klimov@jetbrains.com}\\
\IEEEauthorblockN{Pouria Derakhshanfar}
\IEEEauthorblockA{JetBrains Research\\The Netherlands\\
pouria.derakhshanfar@jetbrains.com}\\
\and
\IEEEauthorblockN{Muhammad Umair Ahmed}
\IEEEauthorblockA{Bilkent University\\Türkiye\\
umair.ahmed@bilkent.edu.tr}\\
\IEEEauthorblockN{Eray Tüzün}
\IEEEauthorblockA{Bilkent University\\Türkiye\\
eraytuzun@cs.bilkent.edu.tr}
\and
\IEEEauthorblockN{Nikolai Sviridov}
\IEEEauthorblockA{JetBrains Research\\Serbia\\
nikolai.sviridov@jetbrains.com}\\
\IEEEauthorblockN{Vladimir Kovalenko}
\IEEEauthorblockA{JetBrains Research\\The Netherlands\\
vladimir.kovalenko@jetbrains.com}

}

\begin{document}

\maketitle

\begin{abstract}
Bus factor (BF) is a metric that tracks knowledge distribution in a project. It is the minimal number of engineers that have to leave for a project to stall. Despite the fact that there are several algorithms for calculating the bus factor, only a few tools allow easy calculation of bus factor and convenient analysis of results for projects hosted on Git-based providers.

We introduce Bus Factor Explorer, a web application that provides an interface and an API to compute, export, and explore the Bus Factor metric via treemap visualization, simulation mode, and chart editor. It supports repositories hosted on GitHub and enables functionality to search repositories in the interface and process many repositories at the same time. Our tool allows users to identify the files and subsystems at risk of stalling in the event of developer turnover by analyzing the VCS history. 

The application and its source code are publicly available on GitHub at \url{https://github.com/JetBrains-Research/bus-factor-explorer}. The demonstration video can be found on YouTube:  \url{https://youtu.be/uIoV79N14z8}
\end{abstract}


\section{Introduction}\label{sec:intro}

Collaborative software development implies the division of work between team members. This can lead to uneven distribution of knowledge among team members. As a result, the departure of a small group of engineers can result in loss of expertise about parts of the project.

This risk can be reduced by assessing the level of distribution of knowledge among team members. 
One way to do this is to use the \emph{bus factor} metric. 
As with other metrics \cite{Unwin2020Why}, clearly visualizing BF of projects is important to help end users in understanding the assessment. 
Thereby, various tools for BF visualization were introduced in previous studies~\cite{almarimi2021csdetector, gitana-busfactor, avelino2016}. 
However, these tools are not always straightforward to use: tool users need to manually clone the target repository and then run additional scripts to analyze the VCS history.

This paper introduces a new, user-friendly, tool called \tool, for analyzing and visualizing bus factor information. 
With our tool, practitioners can get BF information on any project available on GitHub in a few clicks. 
In addition, we designed \tool to be used by developers \emph{and researchers} to visualize BF and create tools based on the bus factor. 
Our tool has two major advantages over existing ones. The first is the flexibility in visualization and accessing data for further analysis. The second is a simulation mode that allows the user to see how the potential departure of contributors impacts the knowledge distribution. 
\tool visualizes how the departure affects contribution scores and bus factor for the whole repository and its folders and files. 


We evaluate the effectiveness of \tool by analyzing \datasetSize popular GitHub repositories. The results show that our tool is capable of analyzing large repositories. For example, our tool can analyze 
12,000 commits in 25 seconds.
\section{Related work}
The research community has created several BF estimation algorithms based on VCS history~\cite{zazworka2010developers, rigby2016quantifying, difficultyOfTruckFactor, cosentino, avelino2016, 9793985}.

In most cases, the estimation consists of several steps.
First, the algorithm mines VCS history to assess the amount of knowledge each engineer has for each file.
Then, for each file, the algorithm produces a list of engineers who are experts in it, so that the departure of all engineers from the list leads to abandonment of the file.
Finally, the algorithm produces the smallest set of engineers such that if all of them leave the project, more than a certain share of the files will be abandoned.
The size of this set is considered the BF of the project.

There only exist a few BF analysis tools with support for visualization or data export~\cite{almarimi2021csdetector, gitana-busfactor, avelino2016}. 
All of them require additional manual steps to prepare the working environment: the users need to clone and process the repository (e.g., scan all commits and store them into files or database using provided scripts) before they can explore the BF data.

\section{Approach}
\subsection{User scenarios}
\textbf{Software development team:} team members can use this tool to improve their software development process by analyzing bus factor data.
They can use the provided treemap view or build their own chart to analyze the distribution of knowledge.

\textbf{Researcher:} \tool is a good starting point for researchers to create new BF calculation algorithms or tools based on the bus factor metric. 
Researchers can calculate the bus factor using the built-in algorithm and export it in CSV and JSON formats for further exploration.
They can also explore the data using the interactive chart editor.

\subsection{Bus Factor Calculation}
Our bus factor calculation algorithm is based on the study by Jabrayilzade et al.~\cite{9793985}, excluding meetings and reviews data.
We select this algorithm, because, in contrast with others, it is based on the Degree of Authorship (DOA) formula that assumes that knowledge from a contribution decays exponentially and halves every five months. 
It is stated \cite{9793985} that this approach yields better estimates than the popular formula by Avelino et al~\cite{avelino2016}.
The top contributing authors are removed iteratively until the current engineers' knowledge covers less than half of the files. 
The number of removed engineers is the bus factor.

\subsection{Visualization}
\begin{figure}[h]
\includegraphics[clip,width=\columnwidth]{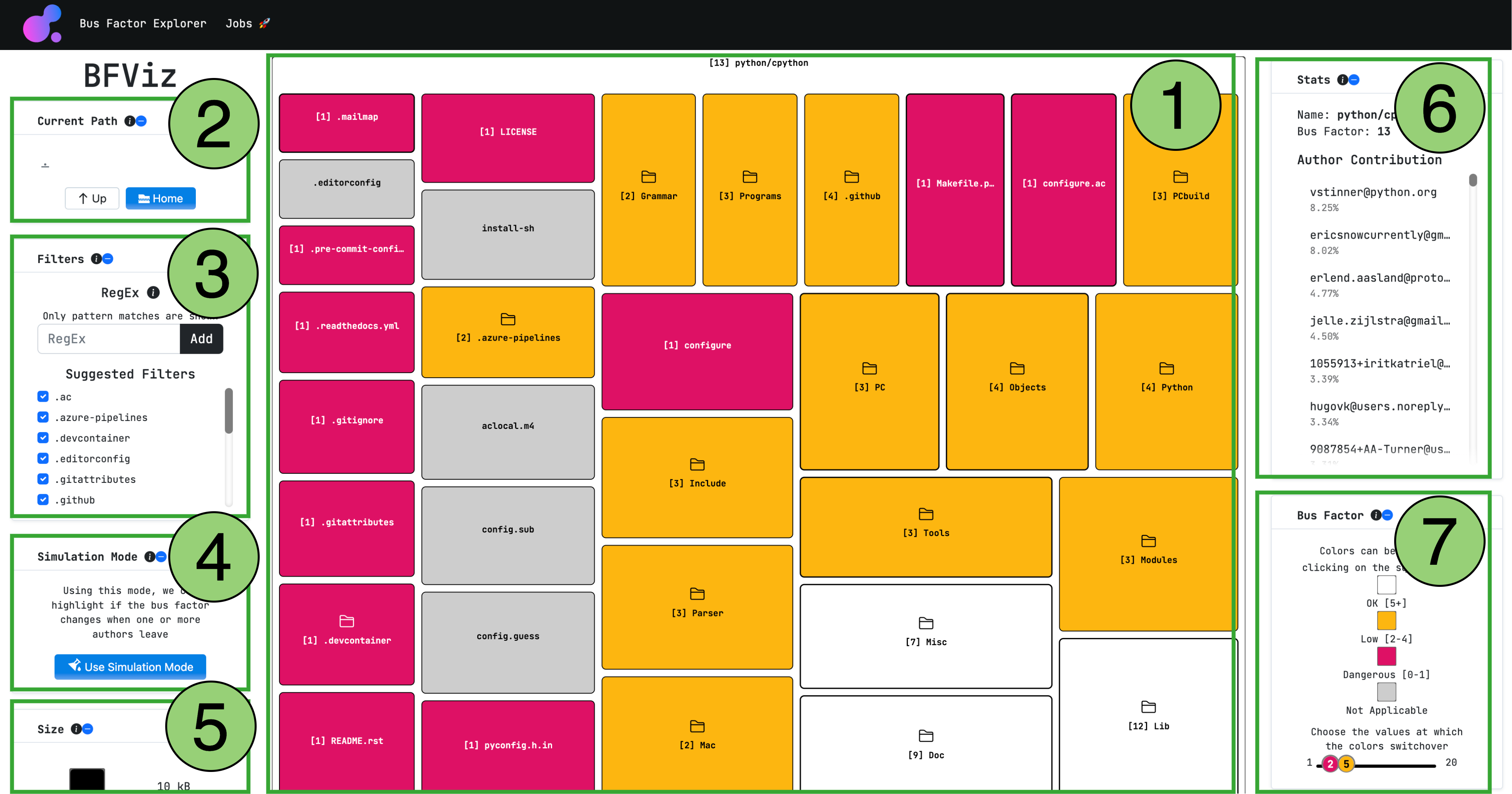}
\caption{Treemap report for the cpython repository}
\label{fig:bfvis}
\end{figure}

Figure \ref{fig:bfvis} depicts how \tool visualizes the bus factor.
The visualization is in the form of a treemap (1 in Figure \ref{fig:bfvis}) which shows the contents (files and folders) of a parent folder.
Each tile on the tree map represents a child node.
We use D3.js\footnote{D3.js: \url{https://d3js.org/}} to implement the treemap. 

The layout is generated using the \texttt{squarify} method from D3.js.
In our tool, the size of each tile represents the size (in bytes) of the corresponding node. 
For this representation, we use the logarithmic scale which is essential as the variance in byte size of the nodes in most projects is very high. 
Without normalization, smaller nodes (e.g., a few bytes) would appear minuscule compared to their larger siblings (e.g., several MB in size).
The tiles are sorted in ascending order with respect to their size in bytes.

The tiles are interactive. 
On hover, the full file name and the bus factor value are displayed. 
A click on a folder node centers the view on its contents. 
The application also updates the view to show author contribution statistics for the clicked node (5 in Figure \ref{fig:bfvis}).
The color of each tile in the treemap is defined by its bus factor. 
There are 4 categories of bus factor values (\ref{fig:bfvis}): \textit{Not Applicable}, \textit{Dangerous}, \textit{Low}, and \textit{OK}.
These ranges and the colors are user-configurable (6 in Figure \ref{fig:bfvis}).

\subsection{Simulation Mode}
\begin{figure}[h]
\includegraphics[clip,width=\columnwidth]{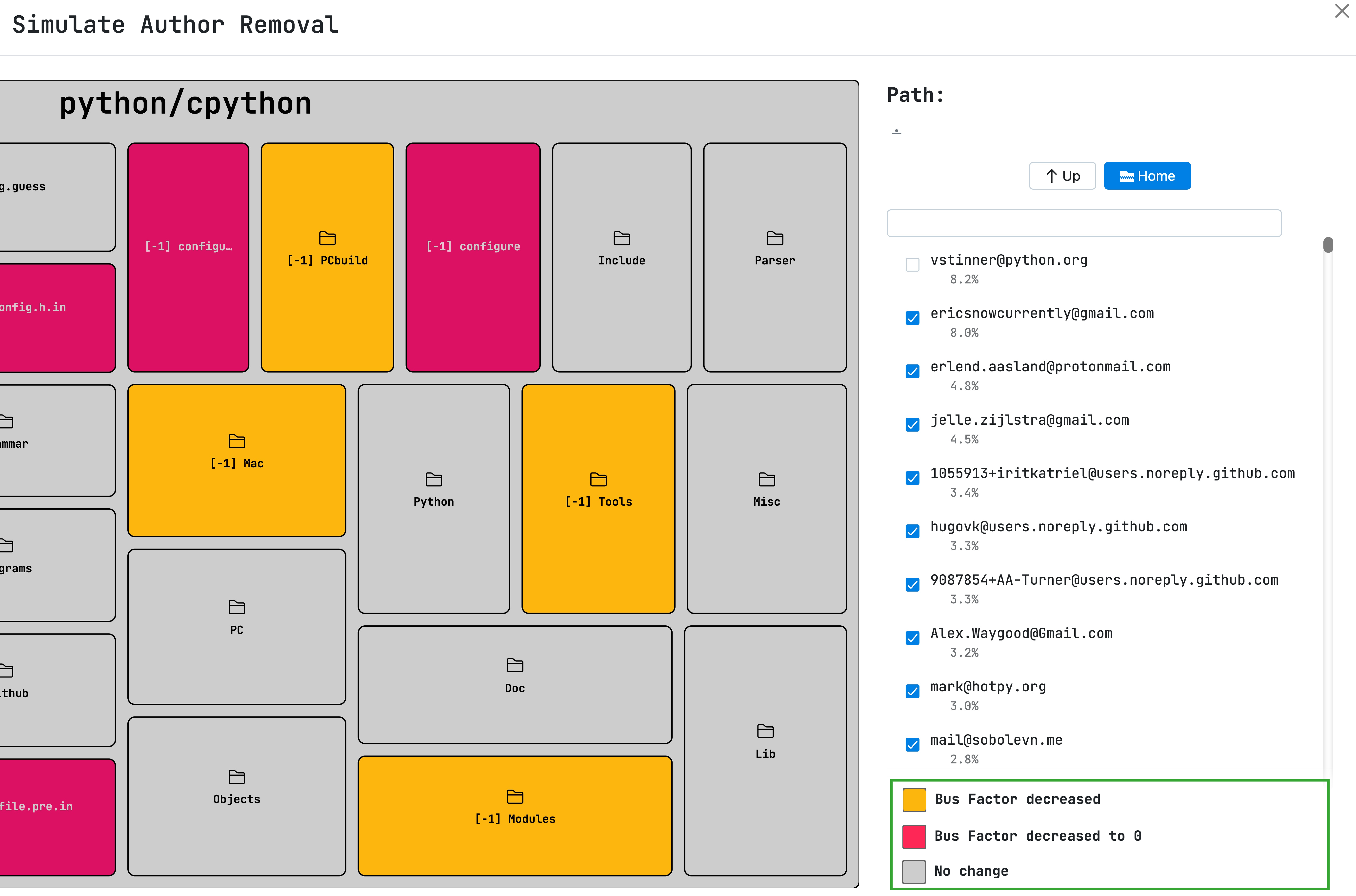}
\caption{Simulation mode for the 
Linux Kernel
repository}
\label{fig:simulation}
\end{figure}

To see the effect of the potential departure of contributors on the bus factor in the project, \tool features a \textit{simulation mode} on top of the treemap view. 
In this mode, the bus factor for the files and folders in a project is recalculated after excluding one or more contributors of the user's choice. 
The original BF information is then compared with the new values calculated after the exclusion.
This relative change in the bus factor is shown for each file and folder (Figure \ref{fig:simulation}).

\section{Implementation}
\begin{figure*}[htp]
\includegraphics[clip,width=\linewidth]{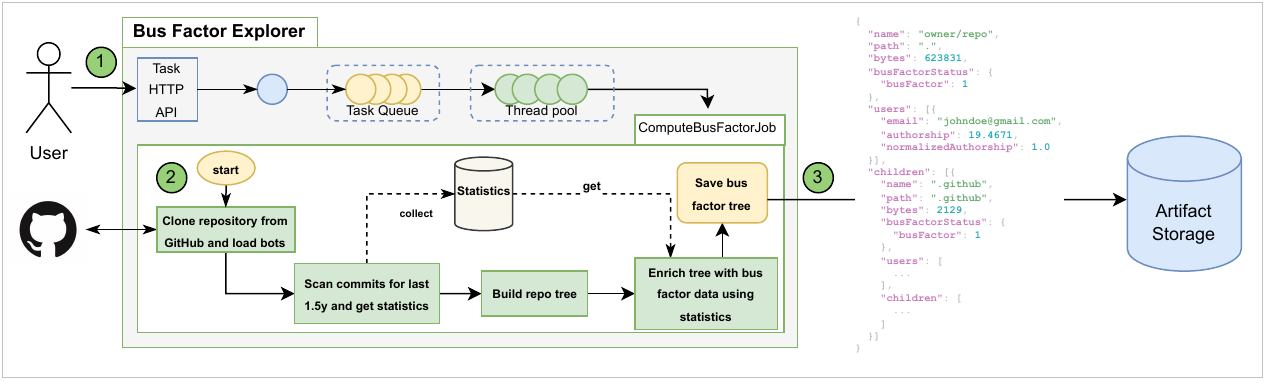}
\caption{An overview of the tool workflow. 
The tool iterates commits for the last 1.5 years since the last commit and collects information about file ownership. Next, it builds file tree for the repository. Then, the tree is enriched by the bus factor data.}
\label{fig:arch}
\end{figure*}

\subsection{Overview and Design}
Figure \ref{fig:arch} represents the main workflow of the tool: computing bus factor for GitHub repositories. 
The user can search for a repository in the main page and submit a \texttt{ComputeBusFactorJob} task (1 in Figure \ref{fig:arch}). 
After that, the service clones the target repository to the working directory (2 in Figure \ref{fig:arch}) and executes the algorithm (described in the previous section) on the main branch of repository to compute bus factor for each file. 

To exclude bots from bus factor analysis, we load all repository contributors with type ``Bot" using GitHub REST API and remove them from authors. 
All artifacts are stored on the file system (3 in Figure \ref{fig:arch}). 
The result of the computation is a file tree of the repository with additional information about contributors and bus factor for each file. 

To ignore inactive contributors, as suggested by  Jabrayilzade et al.~\cite{9793985}, we only process commits and files for the last 1.5 years since the last commit, and mark old files as inactive.

Users can set the GitHub authentication token in the \texttt{GH\_TOKEN} environment variable. 
It can be used to access their private repositories and to improve the API request rate.

The user is notified about all important steps of the analysis via UI notifications. 
Job log is accessible during computation and can be found in the ``Jobs" page. 
As soon as job results are accessible, the target repository appears on the main page. 
Then, by pressing on the repository name, the user can navigate to the visualization page.

\subsection{How to explore Bus Factor data?}
The visualization page contains built-in visualization with a treemap chart (shown in Figure \ref{fig:bfvis}) (labeled 1) and additional panels (2, 3, 4, 5, 6). 

Clicking on folder nodes navigates the view \textit{into} a folder. 
The navigation panel on the top left can be used to move to any folder on the current folder's path. 
The color assigned to each tile, based on its bus factor category, can be changed by adjusting the corresponding color in the color legend panel on the right. 
Additionally, the ranges corresponding to colors can also be modified by the range slider below the color legend.
A list of contributors with their contribution percentages to the current folder is also shown on the right side.
If the current folder has a bus factor of $N$, the top $N$ names are listed.

A few additional panels contain actions to explore data.
\textbf{Simulation Mode}: This interface is activated by pressing the \textit{Use Simulation Mode} button on the Simulation Mode panel. 
This panel includes a secondary treemap and a list of contributors with their contribution percentage to the currently visualized folder and its contents.
Each contributor has a checkbox next to their name.
To view the effect of their departure on the bus factor of the project and its files, the user can uncheck the checkboxes next to contributor names.
This will trigger an update of the simulation treemap.

\textbf{Explore Data}: this panel contains buttons to work with the bus factor data. 
One available option is to download the result in JSON and CSV formats. 
JSON structure is shown on the right side of Figure \ref{fig:arch}. 
Each row of the CSV file is a source file and has the same properties as a JSON and a generated ID and tree path. 
These features can help researchers create other visualization and analysis tools based on the bus factor data.

\begin{figure}[h]
\includegraphics[clip,width=\columnwidth]{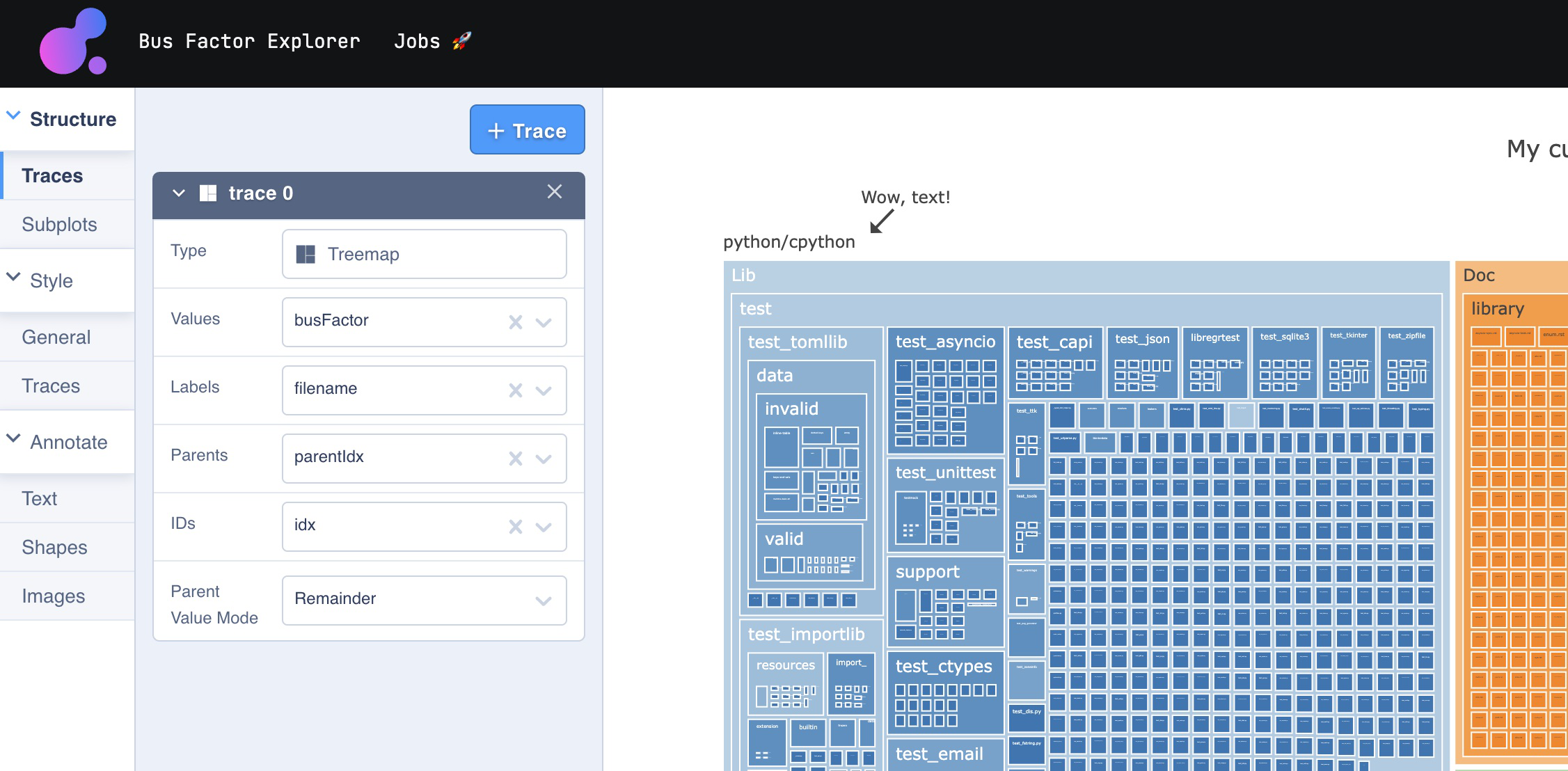}
\caption{Treemap report for cpython repository built with interactive chart editor}
\label{fig:plotly}
\end{figure}

The second option is to use an interactive chart editor, based on the Plotly\footnote{Editor panel for Plotly charts: \url{https://github.com/plotly/react-chart-editor}} chart editor. 
In this case, a CSV file with bus factor data is used as a data source.
Users can build different types of charts, such as scatter, bar, and many others. 
Also, Plotly provides data transformation capabilities and allows to change chart style. 
After each modification, chart settings are stored on disk, so that the same chart is available after reloading the page. 
The chart can be downloaded as \textsc{png} file. A treemap built with the Plotly editor is shown in Figure \ref{fig:plotly}.


\subsection{Distribution}
The tool is distributed as a single Docker image, hosted on GitHub Packages. This makes it possible to start it by a single Docker command locally or on a remote machine. 
A Docker Compose configuration with already analyzed projects is present in the source code.

\section{Evaluation}\label{sec:eval}

\subsection{Perfomance evaluation on real-world projects}

\begin{figure}[h]
\includegraphics[clip,width=\columnwidth]{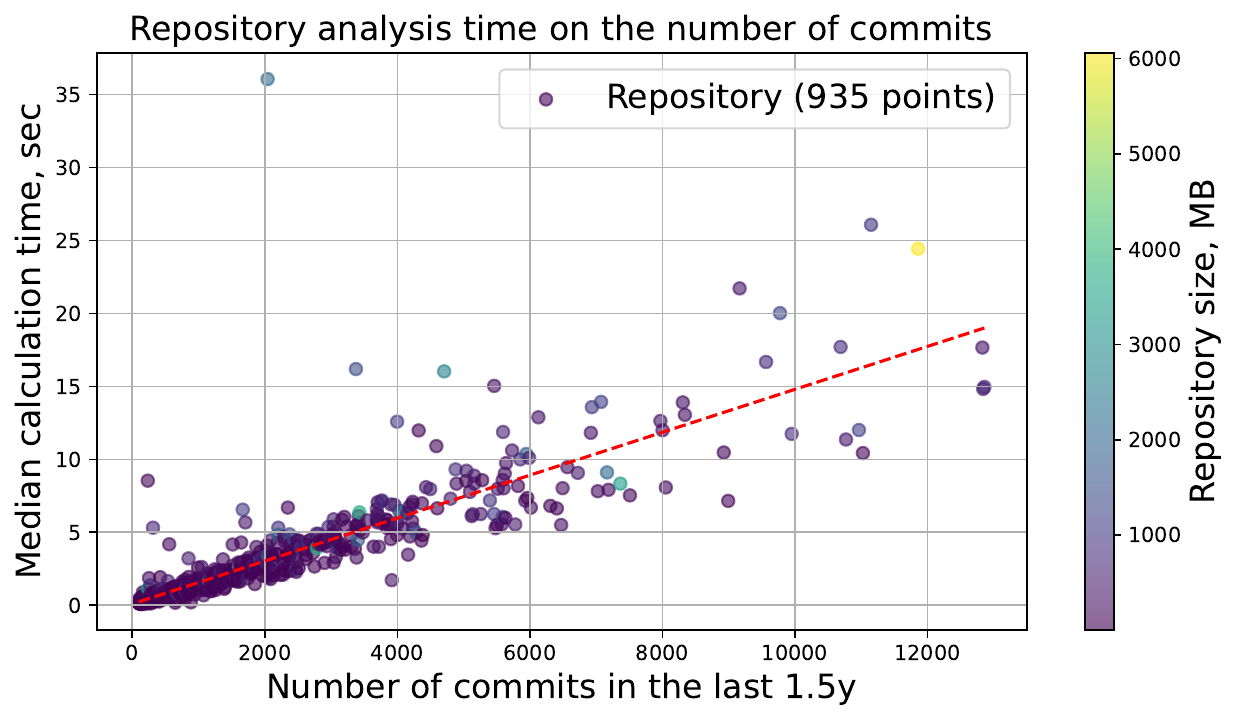}
\caption{Discovered dependency of repository analysis time on the number of commits}
\label{fig:evaluation}
\end{figure}

To evaluate the performance of \tool, we collected a dataset of \datasetSize repositories using the GitHub GraphQL API.\footnote{GraphQL API documentation: \url{https://docs.github.com/en/graphql}} 
We included repositories that have at least 100 commits over the past 1.5 years, are labeled with a language, and occupy up to 10GB on disk (to keep the resource consumption sane). 

For each repository, we calculated the bus factor 10 times using the tool API, and collected information about the execution time. 
We modified the code slightly to avoid cloning the repository on each run. 
Figure \ref{fig:evaluation} demonstrates the dependency of the analysis time on the number of commits. The figure suggests a linear dependency. Table \ref{table:dataset} describes common information about repositories in our dataset and shows information about top 5 languages in it by count of repositories. It should be noted that the algorithm running time did not exceed 36 seconds for projects with a large number of commits. The peak consumption of RAM during the experiment was 1 GB. 
The experiment was carried out on a laptop with an Apple M1 Max processor (10 cores). The data and the code for the evaluation are available in the \textsc{evaluation} directory of \tool repository.

\begin{table}
\caption{Data description}
\label{table:dataset}
\begin{tabular}{lccccc}
\hline
Language & Repos & Commits & Commits & Size, MB & Time, sec \\
& & (total) & (median) & (median) & (median)\\
\hline

\textbf{TypeScript} & 166 & 298,072 & 983 & 74.5 & 1.375 \\
\textbf{JavaScript} & 137 & 139,521 & 394 & 43.5 & 0.512 \\
\textbf{Python} & 128 & 182,847 & 719 & 41.5 & 0.994 \\
\textbf{Go} & 110 & 150,623 & 725 & 33.0 & 0.921 \\
\textbf{C++} & 68 & 119,267 & 780 & 74.2 & 1.137 \\

\hline 
\textbf{Total} & 935 & 1,416,562 & 727 & 60.8 & 1.017 \\ 
\hline

\end{tabular}
\end{table}

\subsection{Feedback and planned validation survey}
As the next step, we are planning an extensive UX study to answer the following research questions:
\textbf{(1)} What specific features, implemented in our tool, users find useful to measure the risk and dependence of a team on individual team members? \textbf{(2)} Are the features implemented in our tool easy to use? \textbf{(3)} Do decision-makers find our tool helpful to measure the risks and dependence of a team on individual team members?


Besides the UX study, we have added a feedback form to the main page of the tool and got positive feedback from colleagues who tried \tool. Some of this feedback was used to improve its UI and UX. 

\section{Conclusion}\label{sec:concl}
In this paper, we introduce \tool, a tool to analyze bus factor information for projects hosted on GitHub.
It is a web-based tool, distributed by a single Docker image, that consists of an interactive UI with data exploration and repository search functionality and a backend responsible for calculating the bus factor from VCS history. Calculation results can be obtained in JSON and CSV formats. Interactive features contain custom treemap visualization, simulation mode, and chart editor for visual data analysis.

Software development teams can use \tool to improve their development process by analyzing the distribution of ownership of components among the developers. 
Research teams can use our tool as a starting point to develop new bus factor calculation algorithms or tools based on the bus factor metric.
To evaluate our tool, we computed the bus factor of \datasetSize popular repositories on GitHub. The results show that our tool has a linear execution time dependency on the count of commits and has good performance. The dataset and results are attached with the source code of the tool.

As further work, we plan to conduct an extensive UX study, improve the bus factor calculation algorithm and introduce the ability to add new Git hosts such as JetBrains Space or GitLab.

\section{Acknowledgements}\label{sec:acknowledgements}
This study was partially supported by The Scientific and Technological Research Council of Turkey (TUBITAK) 3501 program (ProjectNumber:121E584)

\bibliographystyle{IEEEtran}
\bibliography{main}

\end{document}